\documentclass[prd,twocolumn,floatfix]{revtex4}
\usepackage{graphicx, epsfig}
\usepackage{color}
\usepackage{mathrsfs}
\usepackage{bm}
\usepackage{latexsym,amssymb,amsmath,float}

\newcommand{\be}{\begin{equation}}
\newcommand{\ee}{\end{equation}}
\newcommand{\bea}{\begin{eqnarray}}
\newcommand{\eea}{\end{eqnarray}}

\newcommand{\gapp}{\mathrel{\raise.3ex\hbox{$>$}\mkern-14mu \lower0.6ex\hbox{$\sim$}}}
\newcommand{\lapp}{\mathrel{\raise.3ex\hbox{$<$}\mkern-14mu \lower0.6ex\hbox{$\sim$}}}
\def\bbox{{\,\lower0.9pt\vbox{\hrule \hbox{\vrule height 0.2 cm
\hskip 0.2 cm \vrule  height 0.2 cm}\hrule}\,}}

\begin{document}
\title{Modified gravity: living without Birkhoff I. DGP }

\author{De-Chang Dai, Irit Maor and Glenn Starkman}

\affiliation{CERCA, Department of Physics, Case Western
Reserve University, Cleveland, OH~~44106-7079}

%%%%%%%%%%%%%%%%%%%%%%%%%%%%%%%%%%%%%%%%%%%%%%%%%%%
\begin{abstract}
% \widetext
We consider the consequences of the absence of Birkhoff's theorem  in theories of modified gravity.
As an example, we calculate the gravitational force on a test particle due to a spherical mass shell
in the Dvali-Gabadadze-Porrati model (DGP).
We show that unlike in General Relativity, the force depends on the mass distribution.
In particular, the gravitational force within a spherical mass shell depends on the geometric structure of the bulk,
and is likely non-zero.
\end{abstract}
%%%%%%%%%%%%%%%%%%%%%%%%%%%%%%%%%%%%%%%%%%%%%%%%%%
\pacs{???}
\maketitle

\section{Introduction}
\indent
An important tool in astrophysics and cosmology is the ability to reconstruct an object's mass
from measurements of the gravitational force in its vicinity (say, by measuring velocities).
Newtonian gravity's inverse-square law guarantees that the gravitational flux through an enclosed surface is conserved,
giving us a leverage on the mass within that surface; this is Gauss's law.
Even though General Relativity (GR) is a highly non-linear theory,
Birkhoff's Theorem (BT) plays the same role  and allows us to reconstruct
the mass of spherically symmetric configurations.
Two particularly important aspects of BT are that:
(a) outside a spherically symmetric mass distribution,
the gravitational potential (i.e. metric) depends on the distribution of the matter density
only at second order in the potential, due to the role of binding energy as a source of gravity;
and (b) a shell of mass has no effect on the metric in its interior.
Without these properties we presumably could not calculate almost any gravitational fields
without knowing details of the distribution of matter all over the Universe.
Indeed many calculations in gravity would become undoable either in principle or practice,
and many others would become enormously more difficult. In general, modifications to GR tend to violate BT \cite{nz}.

Nevertheless, there has been considerable interest in theories that modify GR and break BT.
The first of two primary phenomenological motivations for interest in such theories
is the observation that at scales above stellar clusters
there is too much gravity to be sourced by the observed matter.
Thus either there is dark matter in such systems, or Newton's Law is not valid in this domain.
The latter possibility, known in a surprisingly successful, albeit  phenomenological form
as Modified Newtonian Dynamics (MOND), explains a range of such systems.
The recent discovery of  covariant theories \cite{Bekenstein:2004ne, Zlosnik:2006zu} which yield
GR, Newton and MOND in appropriate limits has further fueled interest in modified gravity
as an alternative to dark matter.

There is also interest in modified gravity theories as an alternative
explanation to dark energy for the observed accelerated expansion of the universe.
Again the existence of reasonable covariant extensions to the Einstein-Hilbert (EH) action,
such as the DGP model of  Dvali, Gabadadze and Porati \cite{DGP} and $f(R)$ theories \cite{fofR}, has been particularly important.

Theories that violate BT should also lose the benefits afforded by it.
It would be expected that the geometry locally will depend on the detailed distribution of mass in the universe. If this dependence is sufficient to qualitatively affect the behavior of dynamical systems, then we  could be  in a most difficult situation of having alternative gravity theories in which we are fundamentally, or at least practically, unable to make firm predictions for a wide variety of important quantities.

We will not perform an exhaustive calculation of non-Birkhoffian behavior in all modified gravity theories, although that may well be a desirable program of research. Instead we will begin that endeavor by considering a particular example, DGP. We choose DGP both because it is a single theory, with only one free parameter of interest, and because the calculations are readily doable. Other modified theories of gravity in general will not respect BT,
and a similar investigation for other theories of gravity is in progress \cite{future},
with preliminary results that are consistent with our expectations that DGP is a reasonable exemplar.

To cut precipitously to the bottom line, by solving the DGP equations for
concentric spherical shells of dust in a background with accelerated expansion,
we will demonstrate the violations of BT that can be expected in modified gravity theories.
The violation in theories that seek to replace dark matter is likely to render the calculability of cosmology
beyond the homogeneous background significantly more difficult or wholly impossible.
Theories which replace dark energy may be safer, although even there we should still expect
significant effects on observable scales and should not be complacent.

The paper is organized as follows. In section \ref{spa} we first review the background de-Sitter solution in DGP theory, and then solve the metric functions on the brane in the presence of a weak source on the brane. We show that the equations define the metric only up to a function, which we name $g(r)$ . We present or modeling for $g(r)$ in section \ref{gra}, and then consider and estimate the severity of the deviations from BT. We conclude in section \ref{con}.

\section{Space time geometry}
\label{spa}
DGP introduces an infinite flat 5-dimensional space (bulk) in which gravitons move freely,
while Standard Model fields are confined to the 4-dimensional brane world. The 5-dimensional gravity theory retains the EH action, while a 4-dimensional EH action is presumably induced by radiative corrections by the matter on the brane. The action of the theory (other than a possible Gibbons-Hawking term) is \cite{DGP}
\begin{eqnarray}
 S_{(5)} &=& -\frac{M_5^{3}}{16\pi}\int dz ~d^{4}x\sqrt{-g}R-\frac{M_{p}^{2}}{16\pi}\int
    d^{4}x\sqrt{-g^{(4)}}R^{(4)} \nonumber \\
 & & +\int d^{4}x \sqrt{-g^{(4)}}{\cal L}_{m} ~.
\end{eqnarray}
Here $z$ is the extra dimension, $M_5$ is the fundamental five-dimensional Planck mass,
$M_{p}$ is the observed four-dimensional Plank mass, $g$ and $R$ are the 5-dimensional metric and Ricci scalar, and $g^{(4)}$ and $R^{(4)}$ are the induced quantities on the brane. The gravity mass scale in the bulk is taken to be extremely low, $M_5 \sim 10^{-3}$eV. This introduces a new physical scale, the crossover scale, at which gravity becomes 5-dimensional, $r_{0}=M_{p}^{2}/2M_5^{3}$. While Einstein gravity approximately holds on the brane on distances shorter than $r_0$, one expects modifications to gravity at larger distances due to the graviton's free movement in the bulk. Actually, bulk effects leak into the brane on even shorter distances than $r_0$, due to non-liner interactions of the scalar degree of freedom of gravity \cite{DGR_effe_g, DGP_nonlinear}. Taking this into account, the length scale at which gravity is modified, known as the Vainshtein radius, is $r_{*}=(r_{0}^{2}r_{s})^{1/3}$, where $r_{s}$ is Schwarzschild radius of the source.
Since $r_{*}\ll r_{0}$, this may open an observational window to the existence of higher dimensions \cite{DGPst}.

An attractive feature of DGP is that it has a branch of cosmological solutions
in which cosmic acceleration occurs without dark energy \cite{DGR_cos}
as the Hubble function approaches $r_{0}$ at late times.
By choosing $r_{0}^{-1}\sim H_0$, DGP presents an alternative mechanism for the present acceleration.
We follow \cite{DGPst} and solve in DGP for the metric of a spherically symmetric,
static matter source in a de-Sitter background.
We then check how the gravitational field changes when we keep the mass constant but change its distribution.
We show that there may be a gravitational force within a spherical mass shell, depending on the form of an undetermined function, $g(r)$.

We start with the Einstein equations in the bulk:
\begin{equation}
 \frac{G_{AB}}{2r_{0}}+\delta (z-z_0) G^{(4)}_{AB}=
        \frac{8\pi}{M_{p}^{2}}T_{AB} ~.
\end{equation}
Here $G_{AB}$ is the bulk Einstein tensor, $G^{(4)}_{AB}$ is the intrinsic Einstein
tensor on the brane, $T_{AB}$ is energy momentum tensor. We have explicitly specified that $T_{AB}$ vanishes except at the location $z_0$ of the brane in the extra dimension (see below, Eq.~\eqref{TAB}).
Without loss of generality, we can take $z_0=0$.

We wish to find the metric (at least on the brane) induced by a static, spherically symmetric (3-d) weak matter
source on the brane, in a background cosmology which, given observations, we take to be de Sitter-like.
This problem was first addressed by Lue and Starkman \cite{DGPst}.
Following them, we write the  line element as \begin{eqnarray}
ds^{2}&=&N^{2}(r,z)dt^2-A^{2}(r,z)dr^{2}-B^{2}(r,z) \nonumber\\
      & &\times [d\theta^{2} +sin^{2}\theta d\phi^{2}]-dz^{2}~.
\end{eqnarray}
We assume a $Z_{2}$ symmetry across the brane, and that there is no spatial curvature in the brane directions, $B\vert_{z=0}=r$. The resulting bulk Einstein tensor and Einstein equations are presented in \cite{DGPst} (equations (2.7) and (2.8) respectively), and we do not reproduce them here.

\subsection{Vacuum solution}
The metric functions for the background de-Sitter solution \cite{DS}
\begin{eqnarray}
\label{DS1}
 N(r,z)&=&\big(1\mp H|z|\big)\big(1-H^{2}r^{2}\big)^{1/2} \nonumber \\
 A(r,z)&=&\big(1\mp H|z|\big)\big(1-H^{2}r^{2}\big)^{-1/2}\\
 B(r,z)&=&\big(1\mp H|z|\big)r ~, \nonumber
 \end{eqnarray}
with the vacuum energy-momentum tensor given by
\be
\label{TAB}
 T^{A}_{B} = \delta(z)\textrm{diag}\big(
        \rho_H,-P_H, -P_H, -P_H, 0 \big) .
\ee
The energy density and pressure due to the bulk are
\be
\rho_{H}\equiv-P_{H}\equiv\frac{3M_{p}^{2}H}{8\pi r_0} \big(r_{0}H\pm 1\big) ~.
\ee

\subsection{Weak source on the brane}
Adding a gravitational source on the brane, the energy momentum tensor remains of the form  \eqref{TAB}, with
\begin{eqnarray}
 &\rho =\rho_{H}+\rho_g(r) ~,~~~
 P =P_{H}+P_g(r) ~.&
\end{eqnarray}
Here $\rho_{g}(r)$ and $P_g(r)$ are the energy density and pressure due to the distribution of matter on the brane. It is convenient to redefine the metric functions,
\begin{eqnarray}
  N(r,z)&\equiv&1+n(r,z)   \label{N(r,z)} \nonumber \\
  A(r,z)&\equiv&1+a(r,z)\\
  B(r,z)&\equiv&(1+b(r,z))r ~.\nonumber
\end{eqnarray}
It can be seen from Eq.~\eqref{DS1} that to first order in $r$ and $z$, we get $n\approx H^2r\mp H|z|$, and the solution for empty space has $\dot{n}(r,0)=\pm H$ (dots denote $d/dz$). Treating the mass distribution on the brane as a perturbation around the vacuum solution we can write
\begin{equation}
\dot{n}(r,0)=\mp H + g(r) ~,
\label{dotn}
\end{equation}
where $g(r)$ is the correction due to the mass distribution. It is here that we deviate from \cite{DGPst}, which took $g(r)=0$. Even though $g(r)$ is not strictly defined by the equations, \cite{DGPst} and \cite{Lue:2005ya} argue that $g(r)=0$ is essential to recover a well-behaved bulk solution. This has been a matter of some dispute, with some authors \cite{DGPss} proposing alternative Schwarzschild solutions. Our own calculations suggest that the solution of \cite{DGPst} itself requires a small non-zero $g(r)$ to recover the desired bulk solution for self-consistency.

We assume that the mass distribution is of negligible pressure, $P_g(r)\approx 0$, and we define $R_g$ and $G_g$ to be the 3-volume integrals over the mass density and $g(r)$ respectively,
\begin{eqnarray}
 R_{g}(r)&\equiv&\frac{8\pi}{M_{p}^2}\int^{r}_{0}
        \tilde{r}^{2}\rho_{g}(\tilde r)d\tilde r ~,
 \label{R_g} \\
 G_{g}(r) &\equiv& \frac{3}{2} r_{0}^{-1} \int_0^r \tilde{r}^{2}}{g(\tilde r)d\tilde r ~.
 \label{G_g}
\end{eqnarray}
We now want to solve for the metric functions $a$, $b$, and $n$.
Except for $G_{zz}$ and $G_{zr}$, the bulk equations can be satisfied by choosing suitable quadratic terms.
It has been shown that $G_{zr}$ is identically zero when covariant conservation of the matter source
and the brane boundary conditions are considered \cite{DGPst}.
The equations we are left with (shown to first order) are the bulk equation $G_{zz}$
\be
\label{eq:gzz}
 n''+\frac{2n'}{r}-2\big(\frac{a}{r}\big)'=\big(\pm H-g
        \big)\big(\dot{a}+2\dot{b}\big)
 - \big(2\dot{a}+\dot{b}\big)\dot{b}
 \ee
(primes are derivatives with respect to $r$ and dots are derivatives with respect to $z$), and the Einstein equations on the brane:
\begin{eqnarray}
\label{eq:Gij}
 \dot{a}+2\dot{b} &=& r_{0}\big(\frac{2a}{r^2}+\frac{2a'}{r}\big)-
        \frac{r_{0}}{r^{2}}R'_{g}
 - 3H\big(r_{0}H\pm 1\big) \nonumber \\
 2\dot{b} &=& r_{0}\big(\frac{2a}{r^{2}}-\frac{2n'}{r}\big)-
        H\big(3Hr_{0}\pm 2\big) - g \\
 \dot{a}+\dot{b} &=& r_{0}\big(-n''-\frac{n'}{r}+\frac{a'}{r}\big)
        -H\big(3Hr_{0}\pm 2\big) - g \nonumber .
\end{eqnarray}
Eliminating $\dot a$ and $\dot b$ and integrating $dr$ yields
\be
\label{eq:r}
 n'r^{2} + ra+\frac{1}{2}H^{2}r^{3}+G_{g}-R_{g}=0~.
\ee
We define a new quantity,
\be
 f(r) \equiv rn'-\frac{1}{2r}\big(R_{g}-G_g\big)+\big(H^{2}\pm
        \frac{H}{2r_{0}}\big)r^{2} + \frac{gr^{2}}{4r_{0}} ~,
 \ee
and use it to replace $\dot{b}$ and $\dot{a}+2\dot{b}$ in \eqref{eq:gzz} and \eqref{eq:Gij}.
Substituting %(\ref{eq:r}), (\ref{eq:a2b}) and (\ref{eq:b})
into (\ref{eq:gzz}),
multiplying by $r^{2}$ and integrating with respect to r yields a quadratic
equation for $f(r)$.
Imposing the boundary condition that asymptotically the solution approaches the de Sitter background yields
\begin{eqnarray}
 f(r) &=& \frac{r}{8r_{0}^{2}}\left[\sqrt{D}-r\left(3-2r_{0}\left(g \mp H\right)
 	\right)\right] \ ; \\
 D(r) &\equiv&\big(3-2r_{0}\big( g\mp H\big)\big)^{2}r^{2} + 12 g r^2 r_0 \nonumber \\
        &+&16 \frac{r_{0}^{2}}{r}
	\Big[Q +  \frac{1}{2}(R_{g}-G_g)
        + r^{3}\big(2H^{2}\pm \frac{3H}{2r_{0}}\big) \Big] \nonumber\\
Q(r) &\equiv &\mp H \big(r_0G_g-\frac{1}{2}gr^{3} \big)
 -\frac{1}{2}\int_0^r g(\tilde r)g'(\tilde r) \tilde{r}^{3}d\tilde r ~.
\nonumber \end{eqnarray}
With this solution for $f(r)$ we get simple expressions for the metric functions on the brane:
\begin{eqnarray}
\label{eq:gravity}
 n(r,0)&=& n_0 + \int_0^rd\tilde r\Big[\frac{f(\tilde r)}{\tilde r} +
        \frac{1}{2\tilde{r}^2}(R_{g}(\tilde r)-G_g(\tilde r)) \nonumber \\
 &-& \frac{g(\tilde r)\tilde r}{4r_0} -
        \big(H^2\pm\frac{H}{2r_0}\big)\tilde r \Big] \\
\label{eq:a}
 a(r,0)&=& \frac{1}{r}(R_{g}(r)-G_g(r))-\frac{1}{2}H^{2}r^{2}-rn'(r,0) \nonumber \\
 b(r,0)&=& 0 ~. \nonumber
\end{eqnarray}
This solution reduces to that of \cite{DGPst} in the limit $g(r)=0$.
We have generalized \cite{DGPst} by allowing the influence of the bulk to be space-dependent, as one expects.
While $g(r)$ approaches zero far from the source, $G_{g}(r)$ reaches a constant non-zero value.
%Looking at Eq.~\eqref{eq:gravity},
The effect of $G_g$ is to modify $R_g(r)$, and the sign of $G_{g}(r)$ will determine whether the gravitational field will increase (suggesting a gravitational source of greater than actual mass) or decrease. The quantity relevant for the acceleration of a test particle is $n'(r,0)$, given by
\begin{eqnarray}
 n'(r,0)&=& \frac{f(r)}{r} +
        \frac{1}{2r^2}(R_{g}(r)-G_g(r)) \nonumber \\
 &-& \frac{g(r)r}{4r_0} -
        \big(H^2\pm\frac{H}{2r_0}\big)r
        ~.
\end{eqnarray}

\section{Gravitational effects of mass shells}
\label{gra}
We shall see that the effects of matter on the metric are not confined to the region exterior to it. To see this we study the gravitational force exerted by a spherical mass shell in both its interior and exterior with a reasonable model of $g(r)$. We next superpose several concentric shells, and construct a mock model of a galaxy, within a cluster and super-cluster for example.

\subsection{modeling}
Based on dimensional analysis we expect a shell with mass $M_i$ and radius $r_i$ to modify $\dot{n}$ by roughly $\sqrt{2GM_i/r_i^{3}}$. By choosing this dimensional form we recognize $g(r)$ as a correction to the local Hubble scale due to the mass shell. We therefore model $g(r)$ as
\be
g(r)=-\sqrt{2GM_i/r_i^3}\times
\begin{cases}
  1                     & {\rm for}~ r\leq r_i \\
  \big(r_i/r \big)^4    & {\rm for}~ r > r_i ~.
\end{cases}
 \label{g(r)}
\ee
Exterior to the shell, $g(r)$ needs to decay strongly enough to ensure that $G_g(r)$ is finite in the limit $r\to \infty$, which is satisfied with our choice of $r^{-4}$. As can be seen later, the dominant contribution of $g(r)$ comes from the interior region, and the details of the decay external to the shell are less important. We have chosen the sign of $g(r)$ to augment the attractive force of  gravity with an eye to our interest in modifications of gravity
that mimic dark matter.

First consider a test particle external to the shell. For distances much larger than the shell radius,
the only relevant contribution comes from the $G_g$ term of \eqref{eq:gravity}.
Asymptotically $G_g \to 2\sqrt{2GM_{i}r_{i}^{3}/r_{0}^{2}}$, which depends on the crossover scale $r_0$ as well as the shell mass and radius.
Notice that the dependence of $G_g$ on the location of the shell is a positive power;
hence for fixed mass the contribution grows for larger shells.
This shows explicitly that the gravitational force of spherically
symmetric systems depends on the distribution of matter.

Now consider a test particle interior to the spherical shell.
Newtonian gravity predicts that the particle feels no acceleration.
For DGP, this depends on the bulk structure.
Taking $g(r)$ non-zero for $r<r_i$ modifies the gravitational potential
and gives a non-zero acceleration {\it within} the shell.
A local deviation of $\dot{n}(r,0)$ from the DGP-de Sitter value
in the presence of a matter concentration would seem to be expected if,
looking at \eqref{dotn}, we can think of $g(r)$ as a local change in the effective Hubble scale.
Overdensities are expected to create such modifications.
This result is extremely disturbing.
It suggests, for example, that when calculating gravitational forces within a galaxy,
one must consider the effects of the cluster and super-cluster in which it is embedded.

\subsection{Multiple shells}
\begin{figure}%[h]
    \centering
    \includegraphics[width=1.7in,height=1.7in]{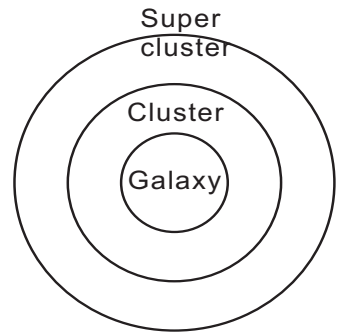}
	    \caption{Mock model of a galactic neighborhood as concentric
            spherical shells of mass.
            }
    \label{ds}
\end{figure}

Notice that $g(r)$ is non-linear in the mass, so one cannot superpose solutions. Concentric shells however, each contributing a $g(r)$ correction to the Hubble scale, is a solvable system. As a mock model of our extended neighborhood we model our galaxy, cluster and supercluster as concentric spherical dust shells, see Fig.~\ref{ds}.
It is self evident that galaxies and clusters are not spherically symmetric and concentric shells, and we have not attempted any data fitting. Nonetheless, the spherically symmetric calculation does give an order of magnitude estimate as to how severe are the deviations between two calculations within the same theory. As we show, the deviations are not negligible. Indeed, another possible way to model the galactic neighborhood is with spheres of mass instead of shells. This will give yet another possible solution within the same theory. As we consider all of these solutions just an estimate to the order of magnitude of the problem, we chose to work with the simple shells model.

\begin{figure}%[h]
    \centering
    \includegraphics[width=3.2in,height=1.7in]{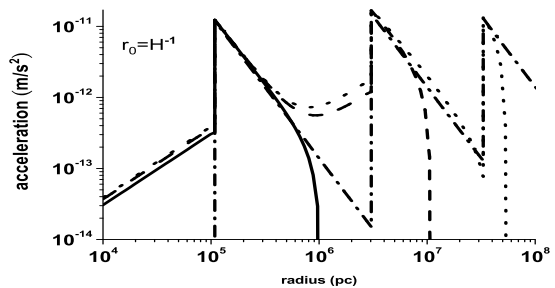}
	    \caption{Gravitational acceleration as a function of radius.
            $g$ (solid) includes only the galaxy as a source,
	        $g+c$ (dashed) adds the cluster, and
            $g+c+sc$ (dotted) adds the super-cluster.
	    The Newtonian acceleration is $g_{n}$ (dot-dashed).
	    The crossover scale is $r_{0}=10^{5}r_{g}=10^{10}$pc$\approx 1/H_0$.
            }
    \label{fig:r0_5rg}
\end{figure}

We first consider DGP as a dark energy alternative, and so take $r_0=1/H$. For the radii of the shells we take $10^{5}$pc,  $3\times10^{6}$pc and  $3\times10^{7}$pc, and $10^{12}M_{\odot} $, $10^{15}M_{\odot} $ and $10^{17}M_{\odot}$ for the masses.
Fig.~\ref{fig:r0_5rg} shows the gravitational acceleration as a function of the distance.
The plot compares the acceleration accounting only for the galaxy (solid),
for the galaxy and cluster (dashed), and for the galaxy, cluster and super-cluster (dotted),
to the Newtonian acceleration (dot-dashed).
Clearly, the accumulating effect of external shells can be non-negligible.
%fig 1
%fig 2
%fig 3
\begin{figure}%[h]
    \centering
    \includegraphics[width=3.2in,height=1.7in]{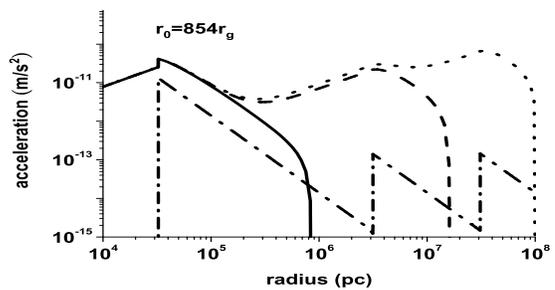}
    \caption{As in Fig.~\ref{fig:r0_5rg}, but with $r_{0}=r_{g}=10^{5}$pc. }
    \label{fig:r0_rg}
\end{figure}
The implication that mass shells can alter the gravitational acceleration interior to them
seems particularly ominous for modified gravity theories that seek to replace dark matter
as an explanation for the anomalous dynamics of galaxies.
Although DGP is {\em not} a candidate MOND theory, we wish to understand, in this calculable theory,
the magnitude of the effects we might expect from BT violations.
We now let our galaxy, cluster and supercluster shells have radii of $3\times10^4$pc,  $3\times10^6$pc and $3\times10^7$pc,
and masses of $10^{11}M_{\odot} $, $10^{13}M_{\odot} $ and $10^{15}M_{\odot}$,
accounting only for the visible matter.
To mimic dark matter, $r_0\simeq 2.6\times10^7$pc is chosen to
boost the acceleration due to the gravity so that the circular velocity of
of a test particle in orbit at $r_g$ is $200$km/s.
As expected, smaller values of $r_{0}$ cause larger modifications to the gravitational force,
as shown in Figure \ref{fig:r0_rg}.
The cluster and supercluster keep the local acceleration approximately constant
from the radius of the galaxy shell outward.
In addition, the internal accelerations to the galaxy shell are substantial.
Clearly ``exterior" mass cannot be ignored.

\section{conclusions}
\label{con}
In this work we have used DGP as an example of the sickness associated with violation of Birkhoff's theorem.
We have generalized the solution of \cite{DGPst},
and shown that (a) even for spherically symmetric systems,
the exact distribution of the matter affects the gravitational force external to the source,
and (b) one {\it cannot neglect} distant gravitational sources,
because even interior to a spherically symmetric mass shell the gravitational force is non-zero.
These effects pose severe conceptual and calculational difficulties.

The problem for DGP may be due to our ignorance of the bulk.
Conceivably a better understanding of how the 4-dimensional metric of the brane is induced
and of bulk effects may change our grim conclusion and answer satisfactorily the issues we have raised. The full understanding of the gravi-scalar degree of freedom near a brane source should hopefully alleviate the ambiguity in the determination of $g(r)$; as argued for example by \cite{DGPst} and \cite{Lue:2005ya}, the gravi-scalar is frozen out, yielding $g(r)=0$, effectively enforcing a Bikrhoff-like vanishing of the gravitational acceleration inside a spherical shell. However, we may find ourselves with a well-determined solution which has a non zero $g(r)$ and thus Birkhoff-violating.

This sickness is not unique to DGP theory:
most theories of modified gravity are likely to suffer the same problem in various degrees.
Since DGP is not a candidate MOND theory,  it is not primarily in their application to DGP {\it per se}
that our results should be viewed, rather as an example of the magnitude of the problem that
is likely to apply to most modified gravity theories.
These issues are the subject of further investigation.
In future work we will also consider deviations from both spherical symmetry and smoothness and how they affect the
gravity in modifications to GR.
A possible conclusion of our work may be that we need Birkhoff's theorem to hold for a theory
to be calculationally tractable.  Nature may, of course, refuse to cooperate.

\acknowledgments
It is a pleasure to thank A.~Lue and P.~Ferreira for very extensive discussions,
as well as F.~Ferrer, I.~Zehavi, T.~Vachaspati and T.~Zlosnik for valuable input.
This work has been supported by DOE and NASA.

\end{document}